\documentclass[twocolumn,amssymb,nobibnotes,aps,prl, reprint,letterpaper,preprintnumbers,amsmath,superscriptaddress,showpacs,floatfix]{revtex4-1}
\usepackage{graphicx,xcolor}
\usepackage{dcolumn}
\usepackage{bm}
\usepackage{graphics}
\usepackage{float}
\usepackage{epsfig}
\usepackage{multirow}
\usepackage{amsmath}
\usepackage{tabularx}
\usepackage{mathrsfs}
\usepackage[pass,paperwidth=8.5in,paperheight=11in]{geometry}

\usepackage{lineno}

\usepackage[linktocpage=true]{hyperref}
\usepackage{hyperref}
\usepackage{comment}
\usepackage{soul}
\usepackage{notes2bib}
\hypersetup{
  pdfnewwindow=true, colorlinks=true,
  linkcolor=blue, anchorcolor=blue,
  citecolor=blue, filecolor=blue,
  menucolor=blue, urlcolor=blue}

\begin{document}

\title{Excitations across the equilibrium and photoinduced `hidden' states of magnetoresistive manganites}

\author{Shiyu Fan}
\email{sfan1@bnl.gov}
\affiliation{National Synchrotron Light Source II, Brookhaven National Laboratory, Upton, NY 11973, USA.}

\author{Feng Jin} 
\affiliation{Anhui Key Laboratory of Condensed Matter at Extreme Conditions, High Magnetic Field Laboratory, and Hefei National Research Center for Physical Sciences at the Microscale, University of Science and Technology of China, Hefei 230026, China}
\affiliation{School of Physics, Hefei University of Technology, Hefei, Anhui 230009, China}

\author{Taehun Kim} 
\affiliation{National Synchrotron Light Source II, Brookhaven National Laboratory, Upton, NY 11973, USA.}

\author{Umesh Kumar} 
\affiliation{Department of Physics and Astronomy, Rutgers University, Piscataway, New Jersey 08854, USA}

\author{Zixun Zhang}
\affiliation{Anhui Key Laboratory of Condensed Matter at Extreme Conditions, High Magnetic Field Laboratory, and Hefei National Research Center for Physical Sciences at the Microscale, University of Science and Technology of China, Hefei 230026, China}

\author{Vivek Bhartiya} 
\affiliation{National Synchrotron Light Source II, Brookhaven National Laboratory, Upton, NY 11973, USA.}

\author{Jiemin Li} 
\affiliation{National Synchrotron Light Source II, Brookhaven National Laboratory, Upton, NY 11973, USA.}

\author{Brandon Yalin} 
\affiliation{National Synchrotron Light Source II, Brookhaven National Laboratory, Upton, NY 11973, USA.}

\author{Yanhong Gu} 
\affiliation{National Synchrotron Light Source II, Brookhaven National Laboratory, Upton, NY 11973, USA.}

\author{Mingqiang Gu} 
\affiliation{Department of Physics, Southern University of Science and Technology, Shenzhen 518055, China}

\author{Wen Hu}
\affiliation{National Synchrotron Light Source II, Brookhaven National Laboratory, Upton, NY 11973, USA.}

\author{Claudio Mazzoli}
\affiliation{National Synchrotron Light Source II, Brookhaven National Laboratory, Upton, NY 11973, USA.}

\author{G. Lawrence Carr}
\affiliation{National Synchrotron Light Source II, Brookhaven National Laboratory, Upton, NY 11973, USA.}

\author{Osor S. Bari\v{s}i\'c}
\affiliation{Department for Research of Materials under Extreme Conditions, Institute of Physics, 10000 Zagreb, Croatia}

\author{Andrey S. Mishchenko}
\affiliation{Department for Research of Materials under Extreme Conditions, Institute of Physics, 10000 Zagreb, Croatia}
\affiliation{RIKEN Center for Emergent Matter Science (CEMS) 2-1 Hirosawa, Wako, Saitama, 351-0198 Japan}

\author{Valentina Bisogni}
\affiliation{National Synchrotron Light Source II, Brookhaven National Laboratory, Upton, NY 11973, USA.}

\author{Sobhit Singh}
\affiliation{Department of Mechanical Engineering, University of Rochester, Rochester, NY 14627, USA}
\affiliation{Materials Science Program, University of Rochester, Rochester, New York 14627, USA}

\author{Wenbin Wu} 
\affiliation{Anhui Key Laboratory of Condensed Matter at Extreme Conditions, High Magnetic Field Laboratory, and Hefei National Research Center for Physical Sciences at the Microscale, University of Science and Technology of China, Hefei 230026, China} 
\affiliation{Institute of Physical Science and Information Technology, Anhui University, Hefei 230601, China}
\affiliation{Collaborative Innovation Center of Advanced Microstructures, Nanjing University, Nanjing 210093, China}

\author{Jonathan Pelliciari}
\email{pelliciari@bnl.gov}
\affiliation{National Synchrotron Light Source II, Brookhaven National Laboratory, Upton, NY 11973, USA.}

\begin{abstract}

“Hidden” phases, generated using ultrafast lasers pulses (few hundreds femtoseconds), with distinct properties than at the thermodynamic equilibrium, are appealing for technologies as they can be long-lived, with a lifetime of hours or weeks, and reversible with temperature sweeping or extra pulses. In this regard, La$_{2/3}$Ca$_{1/3}$MnO$_3$ (LCMO) stands out due to its tunability through epitaxial strain, which can drive the bulk ferromagnetic metal (FMM) into an antiferromagnetic insulator (AFI), and its susceptibility to photo-induced transitions. Indeed, AFI LCMO displays a long-lived photo-induced transition into a putative  `hidden' phase whose exact nature and excitations are still largely unknown.
Here, we combine ultrafast photo-excitation in the near infrared with \textit{in-situ} transport, x-ray absorption (XAS), and Resonant Inelastic X-ray Scattering (RIXS) to investigate the excitations (polarons, phonons, and orbital) of the photo-excited phase of LCMO and contrast them with the thermodynamic phases achieved through strain and temperature. In the thermodynamic regime, we establish the correlation between polarons and transport placing them in the `strong coupling' regime of the Holstein model. Upon photo-excitation of LCMO - AFI, we uncover a long-lived phase characterized by the softening of the polaron excitations, the partial suppression of the Jahn-Teller distortion, and nearly unchanged phonons, showing the emergence of a photo-excited state absent in the equilibrium phase diagram. Finally, by varying temperature, epitaxial strain, and photo-excitation fluence, we construct a polaron phase diagram and identify the key spectroscopic signatures of each phase. Our laser-RIXS approach establishes a versatile platform for exploring photo-induced ‘hidden’ phases in quantum materials in non-stroboscopic conditions.

\end{abstract}

\maketitle

\section{Introduction}

Ultrafast laser pulses provide a powerful route to coherently excite materials into a non-equilibrium state, enabling phenomena such as metal-to-insulator transitions,  multiferroicity, and enhanced superconductivity \cite{McLeod2019,Li2019,Nova2019,Zhang2016,Teitelbaum2019,Gao2022,Dean2016,Fausti2011,Zong2019,Mankowsky2014}. While most of these excited states are transient, lasting only few femto- to picoseconds \cite{Beaud2014,Esposito2017,Stoica2022}, long-lived phases can be created if the ultrafast laser pulse provides sufficient energy to cross/modify the free-energy barrier, driving the system into a metastable local minimum in the free energy landscape \cite{Zhang2016,Teitelbaum2019,Sun2020,Morrison2014}. These `hidden' phases are thermodynamically inaccessible and can persist for hours or longer due to the energy barrier that prevents relaxation to the ground state \cite{Zhang2016,Teitelbaum2019,Li2019,Mitrano2024}. Importantly, such states can often be erased and recovered using additional external stimuli, such as temperature, magnetic field, or laser pulses at different wavelengths, and are therefore amenable for device applications \cite{Zhang2016,Teitelbaum2019}. 
Indeed, the ability to switch between pristine and photo-induced states on demand holds strong potential for ultrafast memory, logic, and quantum information applications based on dynamic control of material properties \cite{Ono2020}.

Quantum materials are susceptible to the emergence of photo-induced long-lived phases due to the interplay of the spin, lattice, charge, and orbital degrees of freedom and proximity of multiple ground states and order parameters \cite{Li2013,Gao2022,Ichikawa2011,Zhang2019}. One exemplar case is the colossal magnetoresistive LCMO, where multiple phases compete due to their close energy scales and intertwined degrees of freedom, offering the opportunity to switch them through static or dynamic perturbations \cite{Jin2020,Hong2020,Zhang2016,McLeod2019}. 

Bulk LCMO displays a low-temperature FMM phase that morphs into a paramagnetic insulating (PMI) phase at room temperature. Epitaxial tensile strain along the $b$ axis induces an AFI phase at low temperature that transitions into the same PMI of the bulk phase above a critical temperature \cite{Jin2020,Hong2020}. Key ingredients to the understanding of the electronic transport in these phases are the Jahn-Teller distortion, the formation of polarons, and electron-phonon coupling  \cite{Millis1996,Kim1996,Yamada1996,Jooss2007,Mertelj2000}.
Moreover, the tendency of LCMO toward phase transitions is highlighted by its switching through photo-excitation leading the AFI phase into a long-lived non-ergodic phase with distinct electronic and magnetic properties \cite{Zhang2016,McLeod2019}. Previous studies have shown that ultrafast excitation can melt the  AFI phase and stabilize the FMM state, often interpreted as a collapse of charge/orbital ordering \cite{Zhang2016,McLeod2019,Teitelbaum2019,Beaud2014}. While the macroscopic switching of resistance and magnetism has been reported in photo-excited epitaxially strained LCMO films \cite{Zhang2016,McLeod2019}, the main spectral features and therefore the microscopic character of the photo-induced state still remain poorly investigated.  

RIXS is a photon-in, photon-out scattering technique with sensitivity to all  the electronic degrees of freedom (spin, charge, orbital, and their coupling to the lattice) \cite{Mitrano2024,Dean2016,Ament2011}. In recent years, progress on the energy resolution and the coupling with other \textit{in-situ} stimuli led to significant advances in the field of condensed matter physics \cite{Bhartiya2025,Mitrano2024}. One of the latest development has been the coupling of RIXS with lasers enabling the study of photo-induced phenomena \cite{Dean2016,Mitrano2024,Hales2023,Pairs2023}. While a significant effort has been put into the development of time-resolved schemes \cite{Dean2016,Hales2023,Pairs2023}, no report are available at this time on the study of photo-induced transition in non-stroboscopic mode targeting metastable phases.

In this work, we combined near infrared femtosecond laser (1030 nm, 1.2 eV, 250 fs) switching with {\it in-situ} transport, x-ray absorption (XAS), and RIXS to study the long-lived transition of LCMO. This multi-pronged approach enables simultaneous access to the transport, the Jahn-Teller distortion, low-energy collective modes, and higher-energy orbital excitations in photo-excited phases. Our central result is that the 1.2 eV photo-excitation on the AFI phase reshapes the Jahn-Teller distortion and polaronic excitations, inducing a long-lived  `hidden' phase microscopically distinct from the equilibrium AFI, FMM, and PMI phases. This state is characterized by \textit{(i)} a metastable reduction of the resistivity; \textit{(ii)} partial suppression of the Jahn-Teller distortion; \textit{(iii)} a substantial softening of the polaronic excitation; and \textit{(iv)} almost unchanged Jahn-Teller-active phonons. These observations coherently point to a partially melted Jahn-Teller polaronic state rather than a direct photo-induced transition to the metallic FMM state. 

This evidence is complemented by data collected by tuning temperature, strain, and excitation fluence, which enable us to construct a manganite polaronic phase diagram and identify an exponential correlation between the polaron excitations and electrical resistivity. This reveals that the polaron excitations of LCMO - FMM, AFI, PMI, and photo-induced phase -  consistently fall within the strongly-coupled Holstein regime. Our results highlight the polaron excitations as a microscopic parameter for tuning electric transport. Within this framework, the photo-induced phase occupies a non-equilibrium region of the polaronic phase diagram that can not be accessed along the equilibrium axes of temperature or strain. More generally, our work proves a novel capability to investigate long-lived photo-induced phases in correlated materials using a high-end scattering technique like RIXS.

This manuscript is divided in the following sections. First, we characterize the electrical transport of LCMO under ultrafast laser switching, and identify a long-lived insulating metastable phase stabilized after photo-excitation (Sec.~I and II). We then compare the microscopic Jahn-Teller distortions, as well as the polaron and phonon signatures, of this photo-induced phase with those of the equilibrium FMM, AFI, and PMI phases (Sec.~III-VI). In these sections, we extract the polaron energies across the equilibrium and photo-induced phases and analyze their relation to the electrical resistivity within a strongly-coupled Holstein framework, and establish how the polaronic renormalization controls transport (Sec.~VII). Afterwards, we discuss that the long-lived photo-induced state corresponds to a distinct Jahn-Teller-polaronic configuration that cannot be reached by temperature or strain alone, and we place these results in the broader context of metastable phases in manganites (Discussion section). Finally, we summarize the main results in the final Summary section.

\subsection{Section I: Evolution of electric transport in LCMO driven by ultrafast photo-excitation}

\begin{figure*}[t!]
\centering
\includegraphics[width=6.5in]{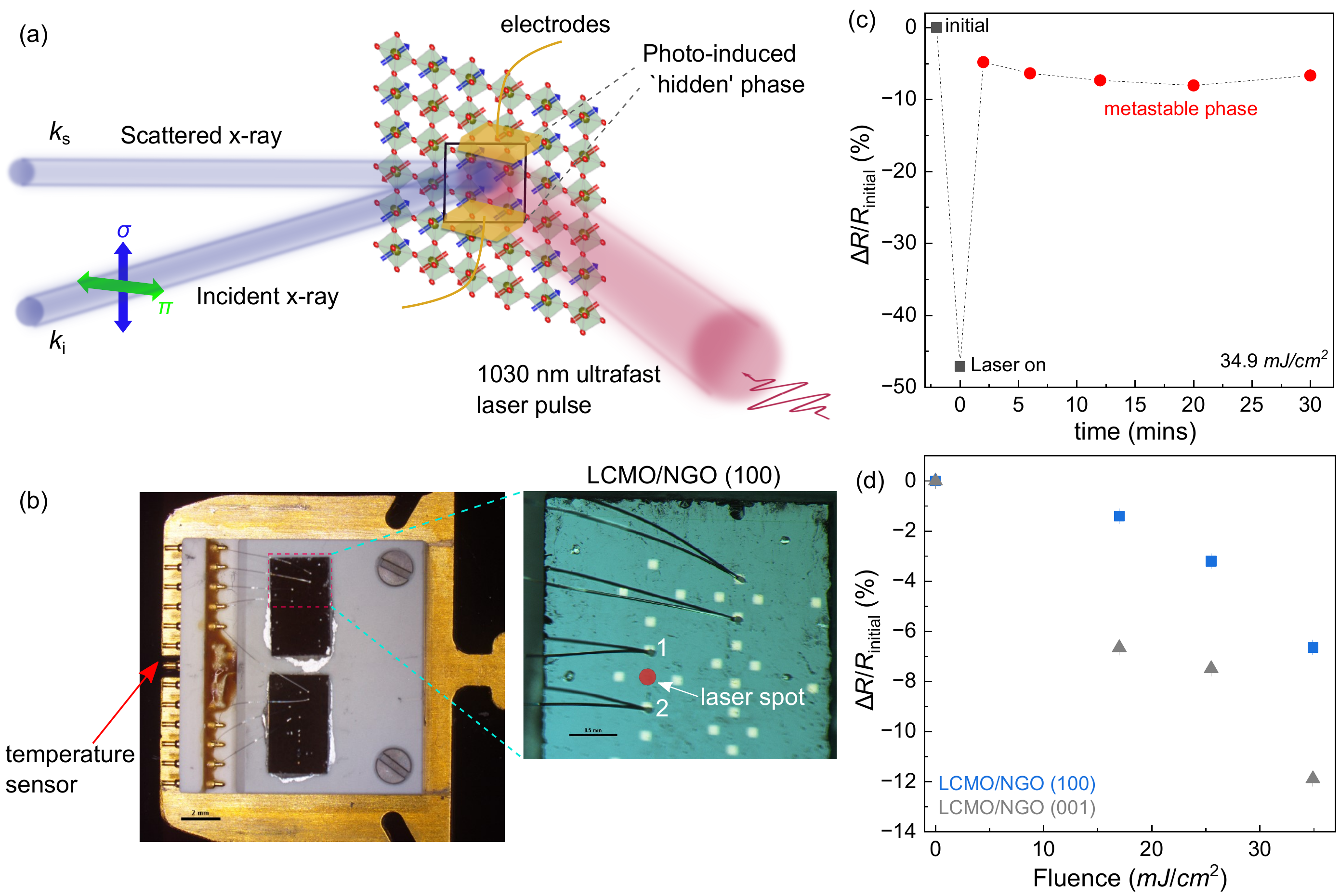}
\caption{ (a) Schematic view of the {\it in-situ} laser-transport-RIXS experimental setup.~The laser pulses are 250 fs and 1030 nm center wavelength. The wave-vector of the incident x-ray ($k_i$) is perpendicular to the ultrafast laser beam. $\sigma$ and $\pi$ refer to the polarization of the incident x-ray as perpendicular or parallel to the scattering plane, respectively. All RIXS spectra are taken with the $\sigma$-polarization. The laser polarization is linearly horizontal, parallel to the scattering plane defined by $k_i$ and $k_s$. The laser beam has a $\sim$ 40 degrees incident angle with respect to the film surface. The RIXS data are collected after photo-excitation in a `non-stroboscopic' manner. (b) Sample holder for the laser-transport-RIXS measurement. The top and bottom films correspond to the AFI LCMO/NGO (100) and LCMO/NGO (001), respectively. A temperature sensor is mounted next to the films to monitor temperature change upon laser irradiation. On the right there is a close-up view of the electrodes deposited on the LCMO/NGO (100) film. The laser spot size is about 120 $\mu$m and the distance between electrodes is 500 $\mu$m. (c) Relative resistance change with respect to the resistance of the initial AFI state ($R_{initial}$) at different recovering times. $\Delta R$ is defined as the resistance difference between the photo-induced and the initial AFI phase. (d) $\Delta R$/$R_{initial}$ as a function of laser fluence of the LCMO - AFI (100) and LCMO - AFI (001) films.}
\label{Structures}
\end{figure*}

LCMO thin films in the FMM and AFI phases were grown on NdGaO$_3$ (NGO) substrates. The films thicknesses are 24 nm. The FMM and AFI phases are characterized by resistivity and magnetization measurements (see Supplementary Information \cite{Supplement}, Fig.~S1). The FMM film is weakly strained and nearly identical to the bulk. The AFI phase is achieved through annealing and exhibits $\sim$ 0.8\% epitaxial tensile strain along the \textit{b}-axis \cite{Jin2020}. The penetration depth of the x-ray and 1030 nm femtosecond laser exceed the film thickness \cite{Zhang2019,Ramesh2024}, ensuring that the entire probed volume is fully excited by the laser [Fig. \ref{Structures}(a)].

To investigate the photo-induced state, we measured the resistance of an LCMO-AFI/NGO (100) film under near infrared (1030 nm, 250 fs) laser excitation by depositing electrodes on the film surface [Fig. \ref{Structures}(b)]. Figure \ref{Structures}(c) shows the relative resistance change over time following laser exposure. The laser was operated at a repetition rate of 1 kHz with a maximum fluence of 34.9 mJ/cm$^2$ for 30 seconds. Upon illumination, the resistance changes by approximately -47\% compared to the pristine state, suggesting that photo-excitation drives the system toward a more conductive state. After the laser is turned off, the resistance partially recovers but stabilizes at a value $\sim$ 7-8\% lower than the initial state [Fig. \ref{Structures}(c)]. Notably, this persistent state remains unchanged even 30 minutes after the laser is turned off, confirming the formation of a long-lived metastable phase whose resistance is distinct from the equilibrium AFI and FMM phases. Heating effects can be ruled out, as all resistance values are measured long after laser exposure. Moreover, given that the laser-irradiated area spans only about one-fourth of the electrode spacing [Fig. \ref{Structures}(b)], the actual change in resistivity would be larger than what is reflected in the measured resistance. These results demonstrate that although photo-excitation at 1030 nm moderately increases conductivity, the photo-induced metastable state still remains within a non-metallic regime. For comparison, we also measured the laser-induced transport in another LCMO/NGO(001) film. A qualitatively similar photo-induced insulating metastable state is observed as well, although the absolute resistance is lower than in LCMO/NGO(100) due to the reduced epitaxial strain [Fig. \ref{Structures}(d)].

In our conditions, the photo-induced metastable phase appears at a laser fluence of $\approx$ 34.9 mJ/cm$^2$, which differs from the reported fluence threshold at around 8 - 12 mJ/cm$^2$ (800 nm) \cite{Zhang2016,McLeod2019,Teitelbaum2019}. While we cannot pinpoint the source of this discrepancy to a single factor, it is possible that it arises from a combination of differences in photo-excitation wavelength, laser pulse duration, and laser pumping protocol (pulse-picked vs. quasi-continuous 1 kHz excitation within a 30-second window) which will be discussed in later sections. 

\subsection{Section II: Ultrafast photo-excitation mechanism in LCMO}

To obtain a more complete view of the multiplet electronic structure of LCMO and better understand the laser-pumping mechanism, we perform RIXS on the LCMO - AFI film at the Mn $L_3$ edge, which is highly sensitive to Mn multiplets and orbital excitations \cite{Ament2011,Jin2025}. Figure \ref{assignment}(a) showcases the RIXS spectra of the LCMO - AFI (100) film at 100 K. We can identify a broad series of peaks ascribable to the multiplets manifold of Mn ions. The peak at 1200 meV correspond to multiplets from  Mn$^{3+}$ $dd$ excitations \cite{Jin2025} which are localized as they are connected to the electronic configuration of Mn and its local geometry. This localization is further evidenced from the RIXS intensity map at the Mn $L_3$ edge, where the 1200 meV excitation shows Raman-like behavior as the incident x-ray energy increases (see Supplementary Information \cite{Supplement}, Section III). The peak at 1600 meV is attributed to an intersite Mn$^{3+}$ to Mn$^{4+}$ charge transfer excitation, consistent with optical spectroscopy \cite{Zhang2016}. The peak below 1 eV can be assigned to a polaron excitation which will be discussed later together with O-$K$ edge RIXS data. Our laser wavelength at 1030 nm (1.2 eV) lies well within the energy window of the Mn $e_g$ manifold, matching the energy of the Jahn-Teller-active Mn$^{3+}$ d$_{3z^2-r^2}$ $\rightarrow$ d$_{x^2-y^2}$ on-site excitation \cite{Jin2025} and overlapping with the lower edge of the Mn$^{3+}$ $\rightarrow$ Mn$^{4+}$ intersite charge-transfer channel [Fig.~\ref{assignment}(b)]. While the on-site transition is typically dipole-forbidden in optical absorption, it becomes active under symmetry-breaking conditions such as O 2$p$ - Mn 3$d$ orbital hybridization or local lattice distortions. Consequently, both excitation pathways may contribute to the laser-induced transitions within the Mn $e_g$ manifold [Fig. \ref{assignment}(b)].

\subsection{Section III: Jahn-Teller effects in the FMM, AFI, and photo-induced phases}

\begin{figure*}[t!]
\centering
\includegraphics[width=6.5in]{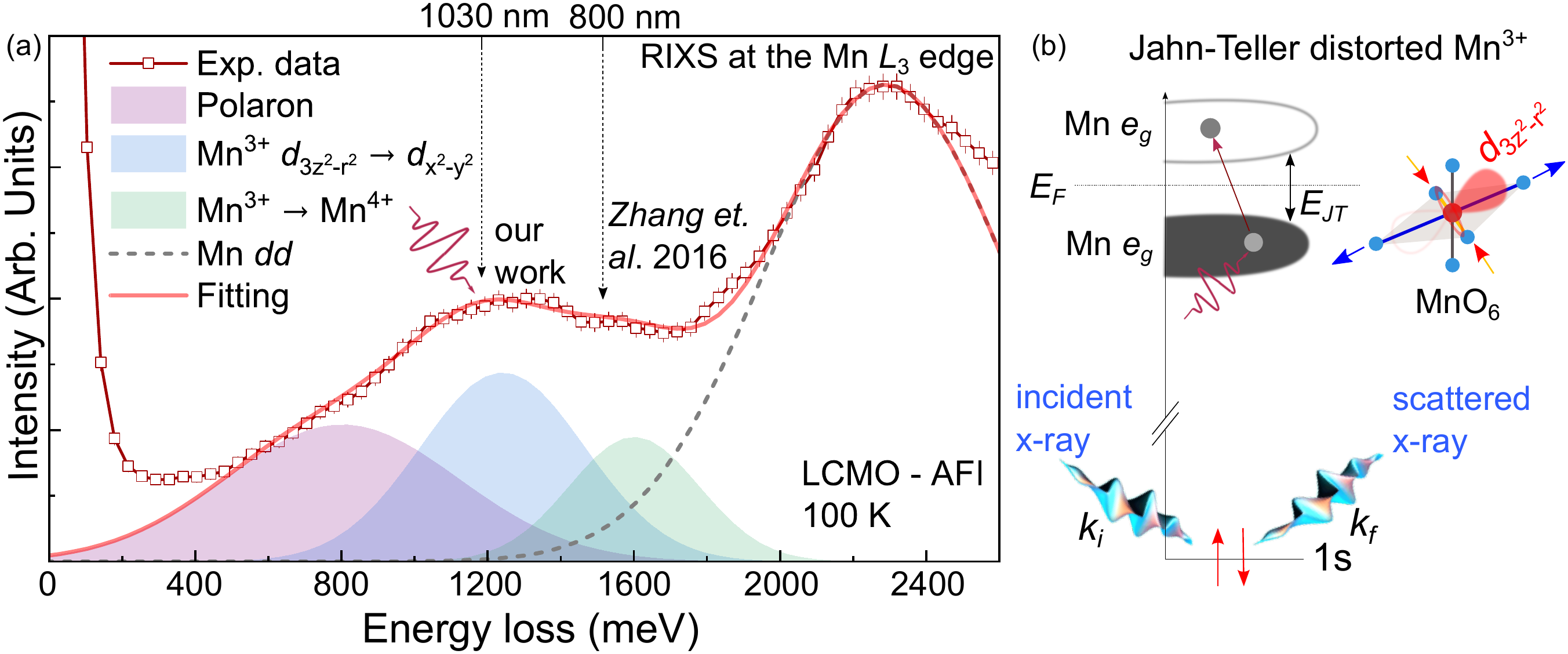}
\caption{(a) RIXS spectrum of the AFI phase of LCMO/NGO (100) at the Mn $L_3$ edge ($E_i$ = 643.2 eV). We indicate the pump photon energies used here (1030 nm, 1.2 eV) and in Refs.  \citenum{McLeod2019,Zhang2016} (800 nm, 1.55 eV). (b) Schematic view of a possible photo-excitation mechanism. While the laser wavelength is on resonance to the Mn$^{3+}$ on-site $d$ - $d$ excitation, the 1.2 - 1.6 eV energy window also overlaps with the Mn$^{3+}$ $\rightarrow$ Mn$^{4+}$ intersite charge-transfer (CT) channel.
}
\label{assignment}
\end{figure*}

\begin{figure*}[t!]
\centering
\includegraphics[width=6.0in]{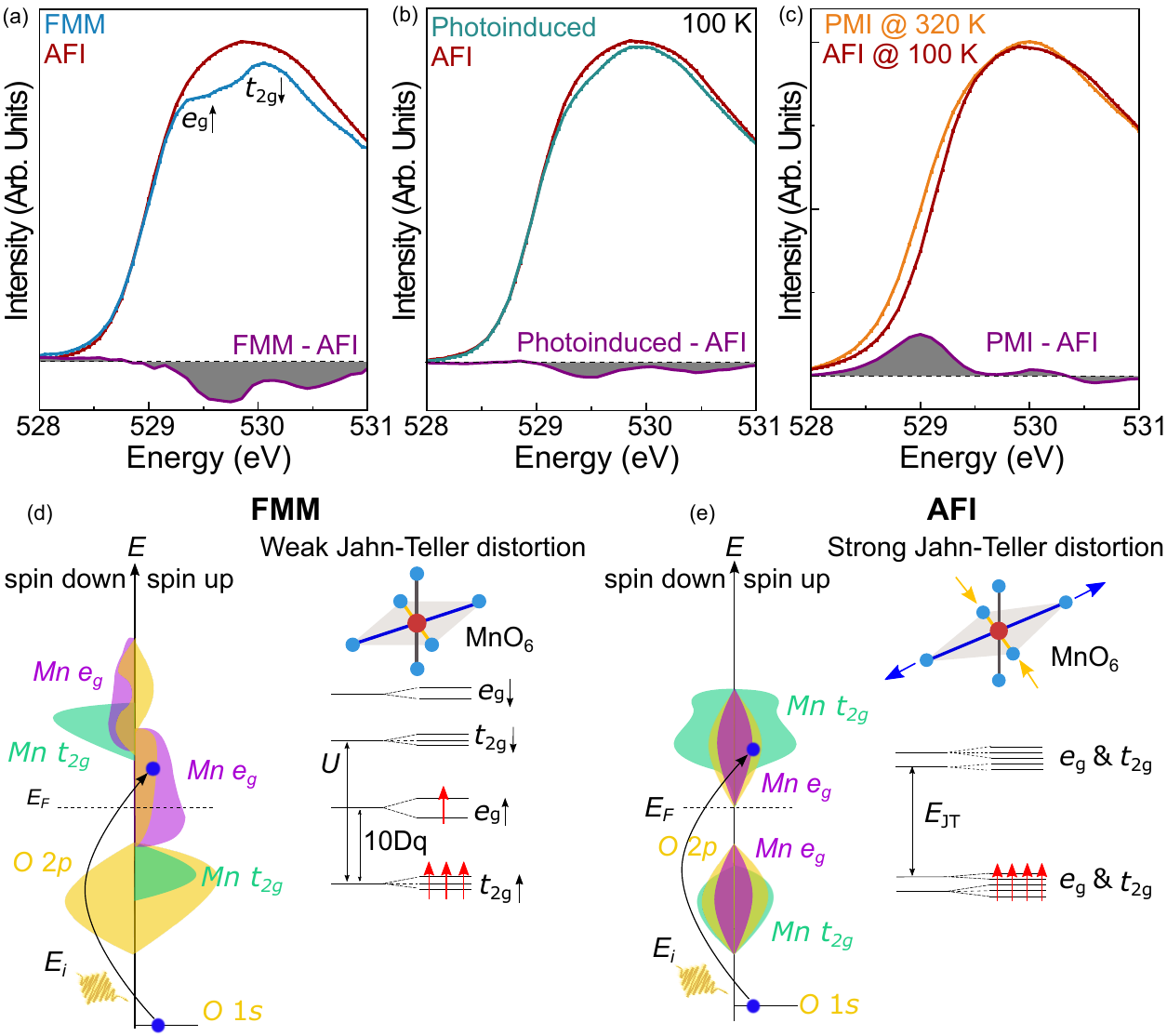}
\caption{(a) O-$K$ edge XAS of the FMM and AFI phases at 100 K. The peaks between 528-531 eV are related to the Mn-3$d$ and O-2$p$ hybridization. (b) O-$K$ edge XAS of the AFI and photo-induced phases at 100 K. (c) O-$K$ edge XAS of the AFI phase at 100 K and the PMI phase at 320 K. Purple curves in (a-c): XAS difference plots of the FMM-AFI, photo-induced-AFI, and PMI-AFI phases, respectively. (d,e) Schematic view of the spin-projected partial density of states (DOS) of the FMM and AFI phases, respectively, along with the O-$K$ XAS process. The DOS of O-2$p$, Mn e$_g$, and Mn t$_{2g}$ states are filled with yellow, purple, and green colors. $E_i$ refers to the incident x-ray energy. The view of the Mn 3$d$ orbital splitting based on the DOS is included for better visualization of the weak and strong $Q_2$ Jahn-Teller distortion for the FMM and AFI phases, respectively. The DOS of the AFI phase is calculated based on an CE-type antiferromagnetic structure \cite{Zhou2011}.} 
\label{XAS}
\end{figure*}

The transport properties of LCMO are affected by the Jahn-Teller distortion, specifically the $Q_2$ mode, where the Mn-O bond length in the basal-plane exhibits out-of-phase displacement [Fig. \ref{assignment}(b)] \cite{McLeod2019,Satpathy1996}. We identify the Jahn-Teller effects of the FMM, AFI, and photo-induced phases of LCMO/NGO (100) through their XAS responses. Figure \ref{XAS}(a) compares XAS at the O-$K$ edge of the FMM and AFI phases at 100 K. A pre-edge peak at $\approx$ 530 eV is associated to the hybridization between the Mn 3$d$ and O 2$p$ orbitals \cite{Valencia2006b,Toulemonde1999}. In the FMM phase, we observe a doublet splitting of the Mn-O hybridized peak while only one broad feature is present in the AFI phase. 
Since XAS is primarily related to the unoccupied density of states (DOS), we can qualitatively discern the XAS lineshape from \textit{ab initio} calculations of the DOS (Supplementary Fig.~13). 
In the FMM phase, the Fermi level is mainly occupied by spin-up Mn e$_g$ itinerant electrons [Fig.~\ref{XAS}(d)]. This is due to a weak Jahn-Teller distortion and gives rise to the half-metallic nature of the FMM phase. The Mn t$_{2g}$ spin-down states are located in the higher conduction band and partially overlaps with the e$_g$ states [Fig.~\ref{XAS}(e)]. Their energy is higher than the Mn e$_{g}$ spin-up states as the on-site Coulomb energy $U$ is stronger than the crystal field splitting $10Dq$. 
Consequently, in the FMM phase, the two well-defined peaks in XAS are associated with the Mn e$_g$ spin-up and t$_{2g}$ spin-down [Fig.~\ref{XAS}(a)]. In the AFI phase, epitaxial tensile strain enhances the $Q_2$ Jahn-Teller distortion, splitting the broad Mn e$_g$ states at the Fermi level and opening up a bandgap. This distortion renormalizes the DOS and leads to a significant overlap of the Mn e$_g$ and Mn t$_{2g}$ levels [Fig.~\ref{XAS}(e)]. As a result, it gives rise to a broad Mn 3$d$ manifolds [Fig.~\ref{XAS}(e)]. Therefore, the broadening of the XAS linewidth and the absence of the e$_g$ - t$_{2g}$ doublet, as shown in Fig. \ref{XAS}(a), are distinctive signatures of the strong $Q_2$ Jahn-Teller distortion in the AFI phase.

Figure \ref{XAS}(b) presents the XAS of the AFI phase before and after photo-excitation. The sample was excited using a 1030 nm (1.2 eV) femtosecond laser with a 1 kHz repetition rate, 250 fs pulse duration, and a fluence of 34.9 mJ/cm$^2$ for 30 seconds - identical to the conditions for the laser-transport measurements. The ultrafast laser is off during the XAS measurement. The spectral weight of the XAS between 529 and 530 eV is noticeably suppressed, whereas it is nearly unaffected around 530 eV. The largest spectral weight suppression is at the e$_g$ peak [$\approx$529.5 eV shown in Fig. \ref{XAS}(a)], indicating that the photo-excitation mainly renormalizes the Mn e$_g$ states while barely affecting the Mn t$_{2g}$ conduction bands [Fig. \ref{XAS}(d,e)]. This increases the e$_g$ - t$_{2g}$ splitting. Indeed, the XAS difference [purple curve in Fig. \ref{XAS}(b)] between the photo-induced and AFI phases closely resembles the lineshape of the XAS difference plot shown in Fig. \ref{XAS}(a). However, the overall effects are weaker compared to the difference between the AFI and FMM phases achieved through epitaxial strain engineering. This indicates that the photo-excitation partially suppresses the $Q_2$ Jahn-Teller distortion as schematized in Fig. \ref{assignment}(b). 

Figure \ref{XAS}(c) compares the XAS of the AFI (100 K) and PMI phases (320 K). The XAS of the PMI phase is broader than the AFI as evident by the increasing spectral weight below 529.5 eV. The XAS difference curve displays a distinct lineshape compared to Fig. \ref{XAS}(a,b). Above 529.5 eV where the e$_g$ and t$_{2g}$ peaks are observed [Fig. \ref{XAS}(a)], the XAS is almost identical for the two phases. The different lineshapes of the XAS difference for the phases at low and high temperatures indicate that the photo-induced and PMI phases are distinct.


\subsection{Section IV: Strain, photo-excitation, and temperature effects on polaron excitations}

\begin{figure*}[t!]
\centering
\includegraphics[width=6.0in]{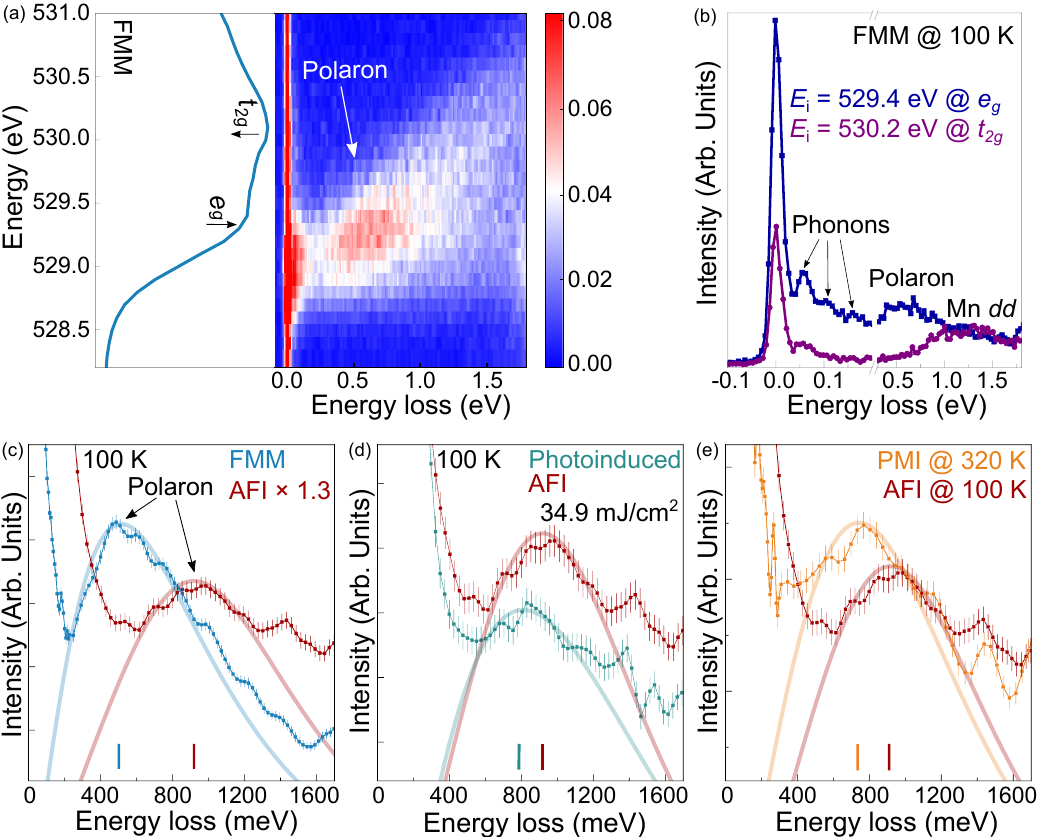}
\caption{(a) RIXS intensity map of the FMM phase as a function of incident photon energy and energy loss. The XAS of FMM at 100 K is included on the left panel. (b) RIXS spectra of the FMM phase at two incident energies: 529.4 and 530.2 eV. The peaks assigned to phonon, polaron, and Mn $dd$ excitations are labeled. (c) RIXS spectra of the FMM and AFI phases at 100 K. The intensity of the RIXS spectra of the AFI phase is enlarged by 1.3 times for better visualization. (d) RIXS spectra of the AFI and photo-induced phases at 100 K. The spectrum of the photo-induced phase is offset along the vertical axis for better visualization. (e) RIXS spectra of the AFI and PMI phase at 100 K and 320 K, respectively. Solid lines in (c-e): Fitting of the polaron peak. Vertical lines at the bottom of each figure indicate the polaron energies. All RIXS spectra in (c-e) are taken with $E_i$ = 529.4 eV.}
\label{RIXS with laser}
\end{figure*}

Figure~\ref{RIXS with laser}(a) shows the O-$K$ edge RIXS of the FMM phase as a function of energy-loss and incident x-ray energy. At this edge, RIXS probes the excitations of the Mn-3$d$ manifolds owing to the strong hybridization between the O-2$p$ and Mn-3$d$ bands [Fig. \ref{XAS}(d,e)] \cite{Fatuzzo2015,Jost2024}. The RIXS map exhibits three main features: a Raman-like phonon around 60\,meV, a pronounced peak at $\sim$\,500 meV, and weaker fluorescence-like features above 1\,eV. 
The high-energy features above 1 eV are better observed at the t$_{2g}$ resonance [Fig. \ref{RIXS with laser}(b)], and are consistent with the excitations of the strongly hybridized Mn 3$d$ - O 2$p$ manifold, with mixed on-site $dd$- and intersite charge-transfer-like character, consistent with Mn $L_3$-edge RIXS and optical conductivity measurements \cite{Quijada1998,Jung1998,Kaplan1996}. The $\sim$ 500 meV mode matches with a polaron response reported in the mid-infrared (MIR) optical studies on LCMO \cite{Hartinger2004,Jung1998,Kim1996} and was also observed in La$_{1.2}$Sr$_{1.8}$Mn$_2$O$_7$ \cite{Jost2024}. We attribute this feature to the excitation of localized carriers interacting with lattice vibrations - a hallmark of polarons induced by strong electron-phonon coupling \cite{Franchini2021,Emin1993,Hartinger2004}. In our RIXS spectra, the “polaron” refers to an excitation arising from the dynamic charge susceptibility associated with locally trapped carriers. The corresponding polaron energy represents an effective energy scale to excite such a trapped carrier into the Mn$^{3+}$ $e_g$ conduction manifold, and is therefore influenced by the Jahn-Teller distortion and electron-phonon coupling. The resonant enhancement of both phonon and polaron intensity at the Mn e$_g$ site [Fig. \ref{RIXS with laser}(b)] highlights the dominant role of Mn e$_g$ electrons in mediating the electron-phonon interaction. Additionally, weak peaks between 100 - 400 meV are observed at the e$_g$ resonance and are ascribed to higher-order phonon harmonics \cite{Jost2024}. 


Figure \ref{RIXS with laser}(c) displays a closer view of the RIXS spectra of the FMM and AFI phases at the Mn e$_{g}$ resonance (529.4 eV). A spectral weight shift from high-to-low energy loss is uncovered across the epitaxial strain-induced insulator-to-metal transition (IMT). The polaron of the FMM phase is more intense and centered at $\approx$\,450\,meV whereas that of the AFI phase is located at $\approx$\,950\,meV. 
To quantify the key features of these polarons we fit the RIXS data. Since the polaron lineshape is highly asymmetric for the FMM phase, we fit the RIXS spectra with an antisymmetric Lorentz function and use the same approach for the AFI \cite{Jost2024}. The detailed fitting procedure is discussed in the Supplementary Information \cite{Supplement}. 


Figure \ref{RIXS with laser}(d) showcases RIXS spectra of the AFI phase before and after photo-excitation. We tune the incident x-ray energy to the e$_g$ resonance where the XAS shows the largest difference [Fig. \ref{XAS}(b)]. After photo-excitation, the polaron energy shifts from $\approx$ 960 meV to 790 meV. While the photo-induced spectral weight redistribution resembles that observed in the epitaxial strain-driven transition [Fig. \ref{RIXS with laser}(c)], the polaron energy remains $\sim$ 300 meV higher than that of the FMM phase. This suggests that the electron-phonon coupling in the photo-induced phase is stronger than in the pristine FMM phase, which is consistent with its more insulating character observed in the laser-transport measurements [Fig. \ref{Structures}(c,d)]. These findings support that the photo-induced phase is not a transition into the metallic FMM phase.

To investigate the polaron across the phase diagram of LCMO as a function of epitaxial strain and temperature, we collected RIXS spectra in the high-temperature PMI phase at 320 K  ($T_C$ = 220 K). Figure \ref{RIXS with laser}(e) compares the RIXS spectra of the AFI and PMI phases. The polaron energy shifts from $\sim$ 960 to 750 meV across the AFI-to-PMI transition. Even though the energy shift of the polaron is comparable for the PMI and the photo-induced phases [Fig. \ref{RIXS with laser}(d)], the photo-induced effects cannot be attributed to thermal heating as the laser was off when collecting RIXS (the acquisition time of a RIXS spectrum is about 3h). Furthermore, the photo-induced phase partially melts the Jahn-Teller distortion [Fig. \ref{XAS}(b)], whereas the PMI phase reflects thermal broadening and disorder but retains the Jahn-Teller distortion as corroborated by XAS [Fig. \ref{XAS}(c)]. This behavior is consistent with the well-known persistence of dynamic Jahn-Teller distortion in magneto-resistive manganites at high temperature \cite{Millis1996,Kim1996,Yamada1996,Zhang2016}.

\subsection{Section V: Phonons of the FMM, AFI, PMI, and Photo-induced phases}

\begin{figure*}[t!]
\centering
\includegraphics[width=6.5in]{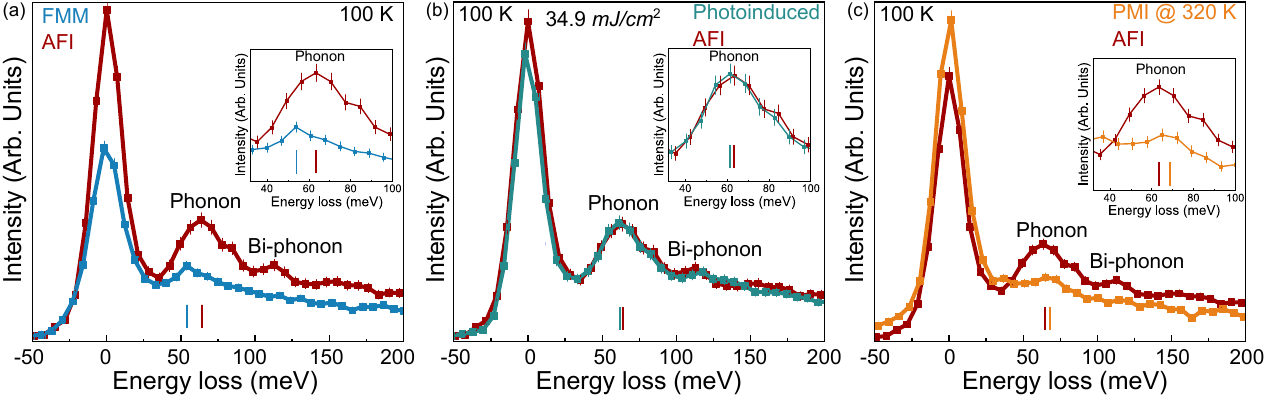}
\caption{ (a) Low-energy RIXS spectra of the FMM and AFI (100) phases at 100 K. (b) Low-energy RIXS spectra of the AFI and photo-induced phases at 100 K. (c) Low-energy RIXS spectra of the AFI and the PMI phases at 100 K and 320 K, respectively. The phonon and bi-phonon are labeled in each figure. The insets show the phonon at $\sim$ 60 meV. Vertical lines at the bottom of each figure indicate the phonon energies. All RIXS spectra are taken with $E_i$ = 529.4 eV.}
\label{RIXS phonon}
\end{figure*}

Owing to the high resolution of RIXS, we can measure phonons in LCMO as a function of temperature, epitaxial strain, and laser fluence. Phonons change dramatically across the epitaxial strain-induced IMT. Figure \ref{RIXS phonon}(a) compares the RIXS spectra below 200 meV for the FMM and AFI phases at 100 K. The phonon at $\approx$ 60 meV is a Jahn-Teller active Mn-O stretching mode \cite{Kim1996,Jost2024}. The FMM phase has a suppression of the phonon intensity, accompanied with about 10 meV red shift of the phonon energy. As the RIXS phonon intensity is proportional to the electron-phonon coupling strength \cite{Jost2024,Braicovich2020}, this suppression and the phonon softening reflect a reduction in coupling due to electronic screening in the metallic phase, consistent with its weak Jahn-Teller distortion and softened polaron energy [Fig. \ref{RIXS with laser}(c)].   

Figure \ref{RIXS phonon}(b) depicts the evolution of the Mn-O stretching mode before and after photo-excitation. In contrast to the prominent softening of the polaron peak [Fig. \ref{RIXS with laser}(d)], the phonon exhibits only a minor red shift and negligible change in intensity compared to the un-pumped AFI state. This suggests that the Jahn-Teller distortion is only partially relaxed, consistent to the reduced e$_g$ - t$_{2g}$ splitting in the XAS of the photo-induced phase [Fig. \ref{XAS}(b)]. Collectively, the photo-induced effects from XAS (Jahn-Teller distortion), RIXS (polaron and phonon), and transport are markedly different from those observed across the epitaxial strain-induced IMT [Fig. \ref{RIXS with laser}(c) and Fig. \ref{RIXS phonon}(a)]. This reinforces the conclusion that the photo-induced and FMM phases are distinct. Within our laser-pumping conditions, our observation contrasts with previous studies using 800 nm laser excitation, where the photo-induced state was sometimes interpreted as resembling the FMM phase \cite{Zhang2016,McLeod2019}. 

The RIXS spectra of the high-temperature PMI phase exhibit a stronger elastic peak, attributed to thermal diffuse scattering, along with suppressed phonon intensity and higher phonon energy compared to the AFI phase. The phonon hardening is more significant near $T_C$ (220 K) which is consistent with the dynamic Jahn-Teller model that links electron-phonon interaction with a double exchange mechanism (see Supplementary Information \cite{Supplement}, Fig.~S11) \cite{Millis1996,Kim1996}. This trend is opposite to the phonon softening seen in the FMM state [Fig. \ref{RIXS phonon}(a)], where the Jahn-Teller distortion is weakened. This reveals that the $Q_2$ Jahn-Teller distortion remains un-relaxed in the PMI phase. The different phonon behavior in the photo-induced and high temperature PMI phases confirms that the photo-induced effects [Fig. \ref{XAS}(b) and Fig. \ref{RIXS with laser}(d)] do not arise from thermal heating, which would otherwise concurrently affect both the phonon and polaron excitations. 

\subsection{Section VI: Fluence dependence of phonons and polaron excitations}
 
\begin{figure}[t!]
\centering
\includegraphics[width=3.5in]{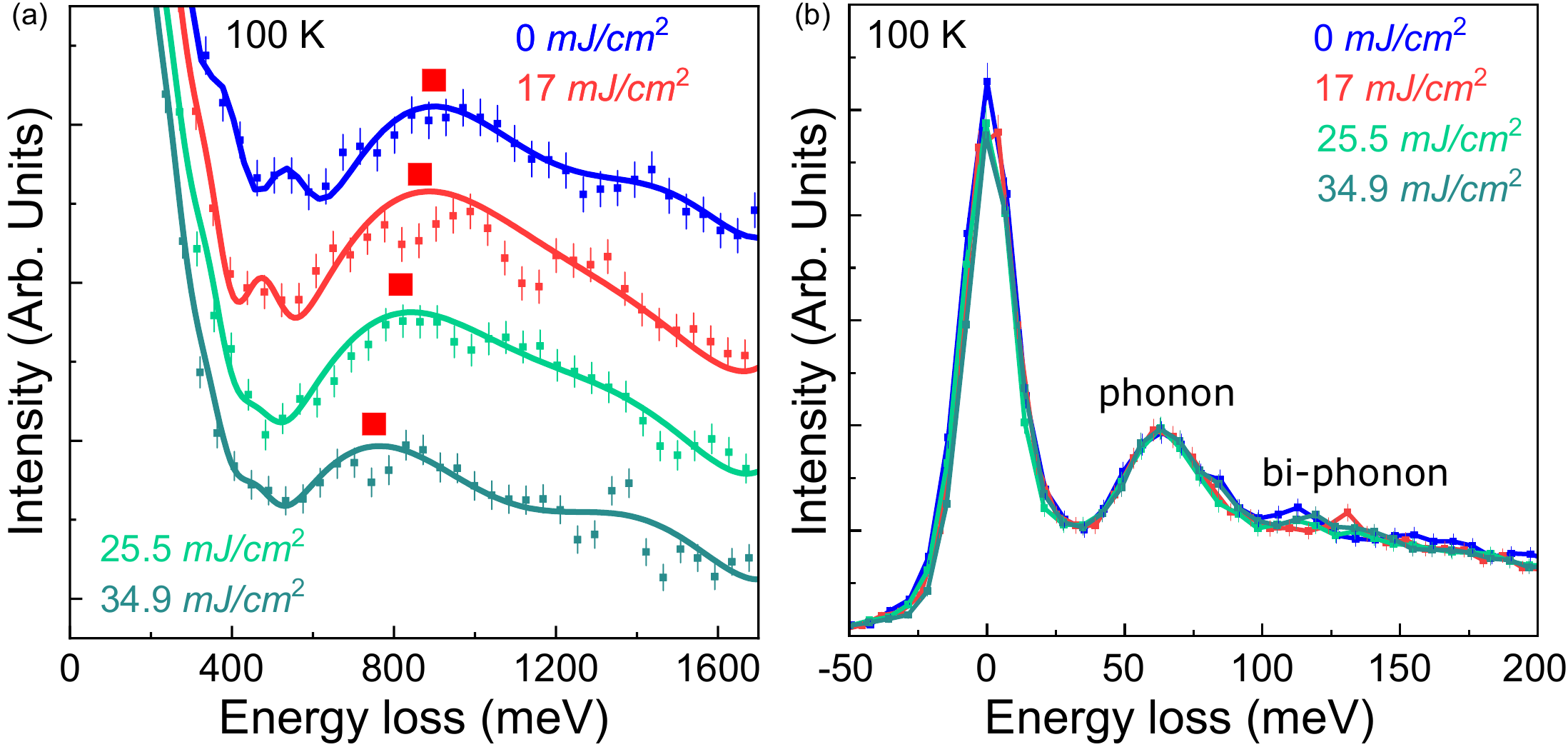}
\caption{ (a) RIXS spectra of LCMO-AFI (100) at energy loss -100 - 1800 meV at different laser fluence. The data are vertically offset for better visualization. (b) RIXS spectra of LCMO-AFI (100) at energy loss from -50 - 200 meV at different laser fluence. The phonon and bi-phonon are labeled. All spectra are taken at E$_i$ = 529.4 eV.}
\label{Fludep}
\end{figure}

To systematically characterize the evolution of the photo-induced phase, we measured the polaron and phonon responses at various laser fluence. Figure \ref{Fludep}(a) shows RIXS spectra of the AFI (100) film as a function of fluence. Using a single-peak fit to the spectral feature between 400 and 1000 meV, the effective polaron energy decreases by about $\sim$ 60 - 180 meV compared to that of the AFI phase as the laser fluence increases. The photo-induced phases at each fluence are long-lived, consistent with the laser-transport data [Fig.~\ref{Structures}(d)]. Remarkably, the 60 meV Mn-O stretching phonon remains nearly unchanged across fluences. 

The noticeable softening of the polaron excitations and constancy of the phonons to photo-excitation indicate that the photo-induced state differs from the FMM. However, with only a micron-scale spatial resolution, the spatial distribution of the photo-induced phase remains unknown with the possibility of phase separation and non-complete switching between the AFI and the photo-induced phases. To better understand this issue, we have prepared another 24 nm LCMO/NGO (100) film which naturally exhibits mixed phase at the equilibrium (see Supplementary Information \cite{Supplement}, Fig.~S2). Unlike the photo-induced phase, this equilibrium mixed-phase sample shows a clear polaron spectral weight enhancement together with a significantly suppressed phonon intensity. Within our spatial and energy resolution, we do not observe a clear two-peak structure of the polaron features in the photo-induced RIXS spectra.~Therefore, this difference highlights that the photo-induced state is unlikely to be composed by a mixed phase of AFI and FMM (see Supplementary Information \cite{Supplement}, Fig.~S2).

\vspace{-0.1in}

\subsection{Section VII: Linking electric transport to polarons} 

\begin{figure}[t!]
\centering
\includegraphics[width=3.5in]{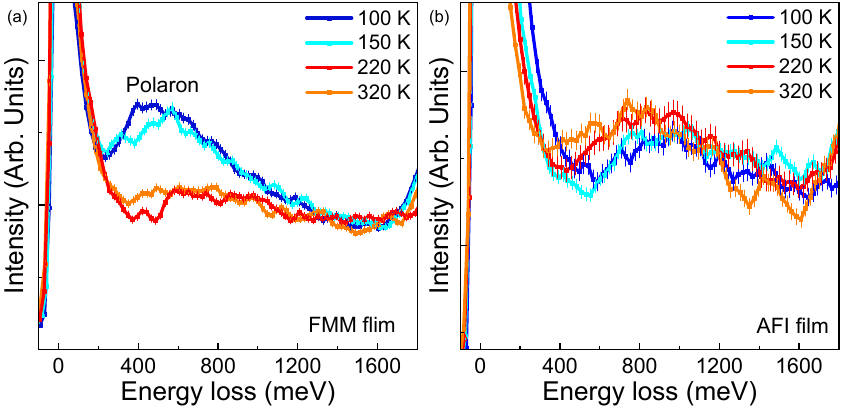}
\caption{(a,b) RIXS spectra of the FMM and AFI at different temperatures, respectively.}
\label{Tdep}
\end{figure}

To relate the polaron excitations to the resistivity and clarify their connection, we measure the evolution of the polaron as a function of temperature. Figure \ref{Tdep}(a,b) displays RIXS spectra of the AFI and FMM (100) films at various temperatures. For the FMM film, the polaron excitations show a spectral weight suppression across the FMM-to-PMI transition at 220 K. The polaron energy has a non-monotonic behavior as the temperature increases: it increases with temperature below T$_C$ (220 K) and decreases with temperature above T$_C$. On the contrary, in the AFI, the polaron energy systematically decreases as the temperature increases, where its spectral weight is less sensitive to the temperature than that of the FMM film. 

\begin{figure*}[t!]
\centering
\includegraphics[width=6.5in]{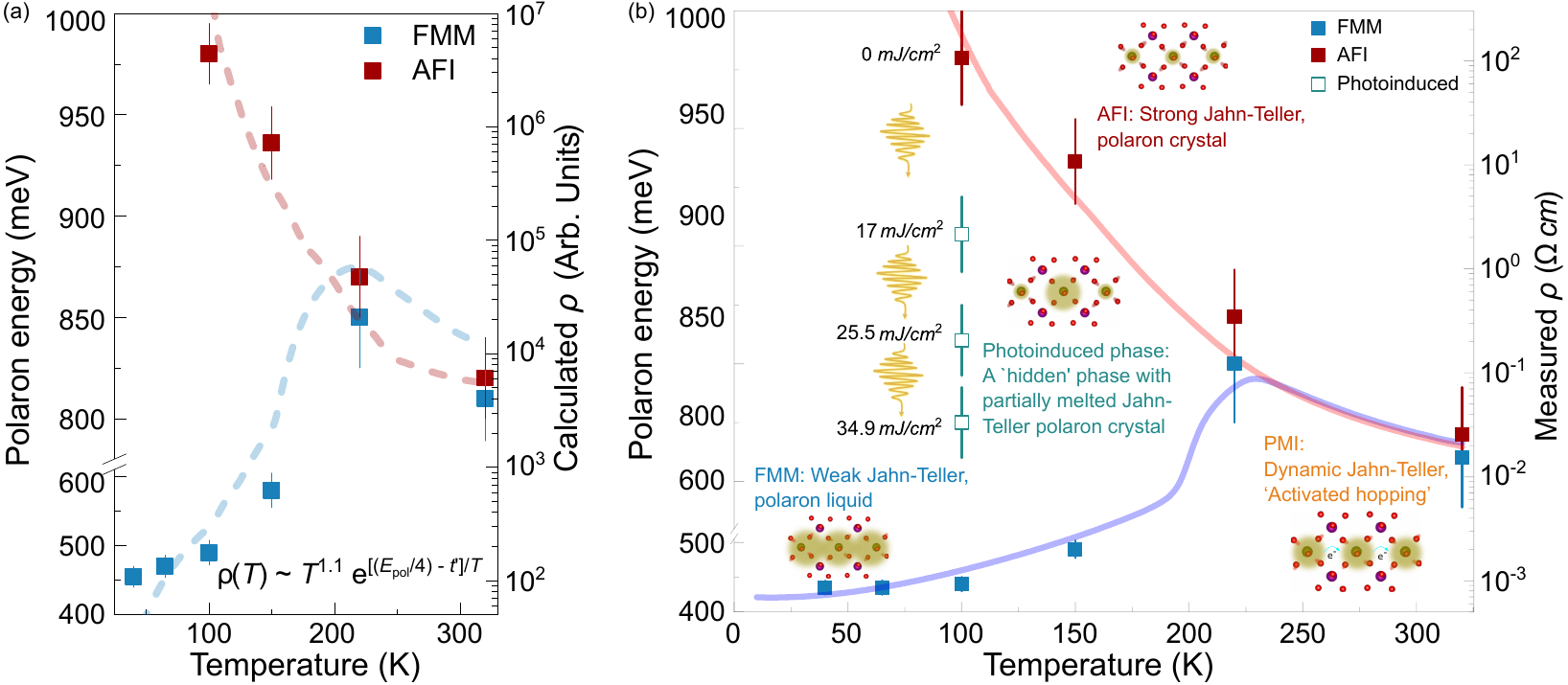}
\caption{(a) Polaron excitation energy of the FMM (blue squares) and AFI (red squares) as a function of temperature from fitting the RIXS spectra using an antisymmetric Lorentz function. Error bars are estimated from multiple fittings. The dashed lines are the calculated electric resistivity using Eq. \ref{Eq1}. We find $t'$ = 1260 K and 1650 K for the FMM and AFI phases, respectively. (b) Polaron excitation energy of the FMM (red squares) and AFI (blue squares) samples as a function of temperature, along with the polaron energy of the photo-induced `hidden' phase at different laser fluence (green squares). The solid lines are the measured resistivity vs. temperature curves of the pristine LCMO-FMM and LCMO-AFI films. The cartoons illustrate the different Jahn-Teller polaronic states corresponding to the AFI, FMM, photoinduced, and PMI phases, respectively.}
\label{Summary of RIXS}
\end{figure*}

We quantify the polaronic excitations energy from fitting of the RIXS data. Figure \ref{Summary of RIXS}(a) shows the extracted polaron energy as a function of temperature for both FMM and AFI. The resistivity changes more than four orders of magnitude with temperature [Fig. \ref{Summary of RIXS}(b)], whereas the polaron energy changes within one order of magnitude. Therefore, a linear change in polaron energy is connected to an exponential change in resistivity. To explain this, we use the strong coupling limit of the Holstein polaron model \cite{Emin1991}. In the `strongly-coupled' small polaron regime ($\lambda_{e-ph}$ $\gg$ 1) of the Holstein polaron, the carrier mobility as a function of temperature ($\mu(T)$) is described by \cite{Mishchenko2015}:

\begin{equation}
\mu(T) = \frac{1}{T^k}\exp({-\frac{E_p/4 - t'}{T}})
\label{Eq1}
\end{equation}
where $E_p$ is the polaron energy. For an Holstein-type polaron, $E_p$ = 2$E_b$, where $E_b$ is the polaron binding energy. $t'$ is the hopping amplitude, $k$ = 1 (1.5) for the adiabatic (non-adiabatic) case. The resistivity $\rho$($T$) is the inverse of the conductivity $\sigma$($T$) = $n\mu$($T$), where we assume for simplicity that the carrier concentration is temperature independent. While this expression originates from a one-dimensional formulation, it captures the essential physics of small polaron tunneling between localized lattice sites and has been shown to be broadly applicable across different dimensionalities. Specifically, the functional form of the temperature-dependent mobility is governed primarily by the electron-phonon coupling strength and the spatial extent of the lattice deformation, rather than the dimensionality per se. This behavior has been confirmed experimentally in two-dimensional manganite thin films \cite{Ziese1998} and through numerous experimental and theoretical studies showing consistent polaron dynamics across 1D and 3D systems \cite{Mishchenko2015,Mishchenko2019,Gunnarsson2003}. Thus, Equation \ref{Eq1} provides a reliable and physically meaningful framework for extracting transport-relevant parameters in the `strongly-coupled' small polaronic regime of LCMO.

Figure \ref{Summary of RIXS}(a) displays the calculated resistivity as a function of temperature. By fitting the temperature dependence of the polaron energy using Eq. \ref{Eq1}, we determine $k$ = 1.1, $t'$ = 1260 and 1650 K for the FMM and AFI phases, respectively. These $t'$ values are physically reasonable when interpreted as effective hopping parameters within the `strongly-coupled' Holstein polaron framework, which are distinct from the bare electronic hopping amplitudes. In the `strong-coupling' regime, the electron hopping is renormalized by the lattice distortion and the polaronic mass enhancement, and thus $t'$ reflects the strong thermally activated mobility of dressed quasi-particles rather than band-like transport \cite{Mishchenko2015}. We note that the extracted hopping amplitudes $t'$ (1260 K and 1650 K) correspond to polaron activation energies on the order of 110-140 meV, which is consistent with the values reported in optical and transport studies of manganites \cite{Emin1993,Ziese1998,Jooss2007}. 

The calculated resistivity curve aligns with the experimental temperature trend of the polaron in all equilibrium phases, demonstrating that the polaron excitations of the FMM, AFI, and PMI phases are all in the `strongly-coupled' regime \cite{Mishchenko2015}. For the photo-induced phase, the polaron also resides in the strong-coupling limit, as its energy falls between those of the FMM and AFI phases, while the phonon energy closely resembles that of the AFI phase. 
According to Eq. \ref{Eq1}, the electric resistivity for a Holstein polaron in the `strongly-coupled' regime depends solely on the polaron excitation energy $E_p$ and the hopping amplitude $t'$. However, since $E_p$ does not explicitly include any temperature-dependent terms \cite{Mishchenko2015}, the strong temperature-dependence of $E_p$ observed in Fig. \ref{Summary of RIXS} arises from changes in the electron-phonon Hamiltonian parameters caused by the $T$-dependent reconstruction of LCMO. Our XAS and RIXS data on phonons indicate that the electron-phonon coupling strength in LCMO is sensitive to the local Jahn-Teller distortion. This suggests that the observed temperature dependence of $E_p$ is strongly associated with the strength of the $Q_2$ Jahn-Teller distortion. 


\section{Discussion}
Figure \ref{Summary of RIXS}(b) summarizes the polaronic states for the AFI, FMM, PMI, and photo-induced phases under epitaxial strain, temperature, and photo-excitation. Even though the polarons in all phases are in the `strongly-coupled' regime, their origins are distinct as their energies are influenced by the strength of the $Q_2$ Jahn-Teller distortion and the amplitude of `activated-hopping' \cite{Mishchenko2015}. From theoretical work \cite{Mishchenko2015}, the `activated hopping' regime occurs when the thermal energy exceeds a quarter of the phonon frequency, resulting in a positive steepness of the polaron mobility ($d\mu/dT > 0$) \cite{Mishchenko2015}. As the polaron mobility is directly proportional to its excitation energy and the strength of electron-phonon coupling \cite{Mishchenko2015}, we can classify different polaronic states of LCMO from the polaron temperature dependence measured by RIXS. The AFI phase has the largest $\lambda_{e-ph}$ as the $Q_2$ Jahn-Teller distortion is enhanced by $b$-axis tensile strain. This cooperative Jahn-Teller distortion (AFI phase) also stabilizes a putative charge- and orbital-ordered structure \cite{Zhou2011,Jin2020,Zhang2016,Teitelbaum2019}. Therefore, the polaron of the AFI phase corresponds to an
ordered array of the localized Jahn-Teller clusters, namely a `polaron-crystal' state \cite{Jooss2007}. This dramatically suppresses the electron hopping between the Mn sites, consistent with its robust insulating behavior (see Supplementary Information \cite{Supplement}, Fig.~S1). For the FMM phase, the polaronic excitation energy is about 500 meV lower than that of the AFI referring to a smaller $\lambda_{e-ph}$ due to the weaker $Q_2$ Jahn-Teller distortion. Furthermore, the polaron energy and resistivity increases with temperature when below 220 K, proving a negative steepness of the polaron mobility [($d\mu/dT < 0$)] [Fig. \ref{Tdep}(a)]. These behaviors indicate
that the polaron of the FMM phase is not in the `activated-hopping' regime (when $d\mu/dT > 0$) as it would require a sizable phonon population unattainable at 100 K. The weak metallicity and `strongly-coupled' polaron suggest that the polaronic state of the FMM phase aligns likely with a `polaron liquid' phase, where the polaron mobility lies between that of a coherent metal and an ordered polaron-crystal insulator \cite{Jost2024}. In the photo-induced phase, the polaron excitation energy systematically softens from the initial AFI phase as the fluence increases. The polaron excitation energy fall between those of the AFI and FMM phases, while the phonon shows minimal changes, indicating the emergence of a new state induced by the partial melting of the Jahn-Teller polaron-crystal state [Fig. \ref{XAS}(b) and Fig. \ref{RIXS phonon}(b)]. The high-temperature PMI phase exhibits a polaron excitation energy similar to the photo-induced phase. However, the origin of the polaronic state differs, as the $Q_2$ Jahn-Teller distortion remains un-relaxed in the PMI phase [Fig. \ref{XAS}(c) and Fig. \ref{RIXS phonon}(c)]. On the other hand, the polaron excitation energy and resistivity decrease with temperature when above 220 K [Fig. \ref{Tdep}(a,b)], consistent with an `activated-hopping' mechanism ($d\mu/dT > 0$) \cite{Mishchenko2015}. The different polaronic states of the FMM, AFI, PMI and photo-induced phases are pictorially summarized in Fig. \ref{Summary of RIXS}(b).

Another factor to explain the photo-induced phase is the role of orbital order. In perovskite manganites such as PCMO and PSMO, the long-lived photo-induced states have been associated with glassy dynamics in the coupled charge-orbital-Jahn-Teller sector, with ultrafast melting of charge/orbital order followed by a slow, domain-limited recovery \cite{Zhou2014,Esposito2018}. The AFI ground state of our LCMO films is likewise stabilized by cooperative Jahn-Teller distortions and possibly orbital order (see Supplementary Information \cite{Supplement}, Fig.~S13). In the photo-induced state, however, we find a substantial softening of the effective polaron excitation energy with fluence, while the 60 meV Jahn-Teller active phonon intensity and energy are only weakly modified and remain less suppressed than in an equilibrium mixed-phase film (Fig.~\ref{Fludep} and see Supplementary Information \cite{Supplement}, Fig.~S2). The 60 meV mode is primarily sensitive to the {\it local} Jahn-Teller distortion amplitude, whereas the polaron excitation energy depends on a combination of factors from both charge and lattice degrees of freedom, including the long-range pattern of the occupied e$_g$ orbitals (orbital order), charge order, and the electronic hopping amplitudes and bandwidth. 
A photo-induced state in which long-range orbital/charge order is strongly disrupted (shorter correlation length, increased domain walls, and/or orbital disorder), whereas the average local Jahn-Teller distortion is only partially reduced, naturally produces a relatively overall unchanged phonon but a much larger shift of the polaron binding energy. This places the photo-induced state between the AFI and FMM phases and is qualitatively consistent with an intermediate orbital configuration that may be more disordered or glassy than in the AFI ground state \cite{Zhou2014,Esposito2018}. At the same time, because our O $K$-edge RIXS measurements do not directly resolve orbital-order superlattice peaks and provide only volume-averaged information, we cannot distinguish unambiguously between a homogeneous phase with reduced orbital order and a more glassy, phase-separated configuration. A definitive characterization of the role of orbital order involved in the photo-induced transition will require future time-resolved resonant x-ray diffraction at the Mn $L$-edge with sensitivity to orbital-order superstructures.

We next comment on the possible microscopic laser-switching mechanism of the photo-induced `hidden' phase, based on electric transport, phonons, polarons, and the evolution of the e$_g$ - t$_{2g}$ splitting associated with Jahn-Teller distortion. According to our DFT calculations, the CE-type AFI phase is charge- and orbital-ordered with the Jahn-Teller active Mn$^{3+}$ and the Jahn-Teller inactive Mn$^{4+}$ sites (see Supplementary Information \cite{Supplement}, Fig.~S13). The Mn$^{3+}$ has a single electron occupying the $d_{3z^2 - r^2}$ orbital which aligns along one of the in-plane diagonal direction of the MnO$_6$ octahedra [Fig. \ref{assignment}(b)] \cite{Jin2020,Satpathy1996,Hashimoto2010}. This orbital occupancy elongates the Mn-O bond along the direction in the Mn-O basal plane, as shown in Fig. \ref{assignment}(b). Consequently, epitaxial uniaxial tensile strain along the $b$-axis enhances the $Q_2$ Jahn-Teller distortion, stabilizing the AFI phase over the FMM phase without altering the doping level. 

Previous studies using an 800 nm pump \cite{Zhang2016, McLeod2019}, resonant with an optically allowed Mn$^{3+}$ $\rightarrow$ Mn$^{4+}$ charge-transfer excitation, reported a dramatic resistivity drop by several orders of magnitude \cite{Zhang2016}. The resulting photo-induced phase was interpreted to be similar in character to the bulk FMM phase. In contrast, our results using a 1030 nm pump suggest a different photo-induced long-lived state, raising the question of whether the excitation mechanism differs. Two possibilities can account for this difference: (i) the 1030 nm laser excites the low-energy tail of the same charge-transfer transition observed at 800 nm, but has a higher fluence threshold; or (ii) the 1030 nm excitation primarily drives an on-site Mn $dd$ orbital transition, specifically from $d_{3z^2 - r^2}$ to $d_{x^2 - y^2}$ based on our Mn $L$-edge measurements and DFT calculations [Fig. \ref{assignment}]. While this on-site excitation is nominally dipole-forbidden, it can become optically active through vibronic coupling and O 2$p$ - Mn 3$d$ hybridization [Fig.~\ref{XAS}]. These two channels likely coexist within the 1.2 - 1.6 eV energy window, consistent with our RIXS analysis [Fig.~\ref{assignment}(a,b)].

In addition to the wavelength, other laser parameters including pulse duration and repetition rate play a role. Our pulses ($\approx$ 250 fs) are substantially longer than the 35 - 50 fs single-shot pulses used in Refs.~\citenum{Zhang2016, McLeod2019}. The longer duration might allow for partial relaxation after the photo-excitation, lowering the transient electronic temperature and thereby increasing the fluence threshold required to cross the non-thermal phase boundary. This may also stabilize a distinct
metastable phase which alters the relaxation pathway within the free-energy landscape \cite{Torre2021,Sun2020}. Similar pulse-duration-dependent selection of driven quantum states has been demonstrated in Floquet and superconducting systems~\cite{Lucchini2022,Budden2021}. Furthermore, our quasi-continuous 1 kHz operation ($\approx$ 3 $\times$ 10$^4$ pulses during a 30 - seconds exposure window) differs from pulse-picked single-shot protocols reported in Refs. \citenum{Zhang2016,Teitelbaum2019}. In this configuration, each pulse can interact with the sample that is still slightly out of equilibrium before fully relaxing to its equilibrium state. Therefore, the successive pulses can partially repopulate or relax pre-existing photo-excited domains, effectively “annealing” the lattice into a higher-entropy configuration by partially melting the orbital order and Jahn-Teller distortion.
This could naturally lead to a weaker or modified `hidden' phase and result in a higher fluence threshold compared to single-shot optical studies \cite{Zhang2016,Teitelbaum2019,McLeod2019}. Single-shot measurements have revealed phases that can be obscured or modified under high-rep-rate, multi-pulse conditions, and related sensitivity of the selected metastable phase to excitation protocol has been reported in 1T-TaS$_2$ and Ta$_2$NiSe$_5$~\cite{Stojchevska2014,Maklar2023,Liu2021}, underscoring the importance of excitation protocol in determining the observed phase and threshold.

Our results show that the nature of the photo-induced phase in LCMO is very sensitive to all these parameters, which opens the possibility for studies based on different excitation protocols. Future laser-RIXS experiments with tunable wavelength, adjustable pulse duration, and capability of pulse-picking, as well as potential time-resolved RIXS measurement will be crucial to separate these effects and understand how different laser conditions reshape the free-energy landscape of manganites.

\section{Summary}
In summary, our findings report on the emergence of a photo-induced ‘hidden’ phase in LCMO and reveal the evolution of the elementary excitations across different electronic phases. We demonstrate that distinct polaronic states can be stabilized through the tuning of epitaxial strain, temperature, and orbital-selective photo-excitation and reveal the connection of the electronic transport with the polaronic excitations of each individual phase. 
Overall, our experimental approach combining ultrafast optical photo-excitation and high-resolution RIXS offers direct spectroscopic access to orbital, phononic, and charge excitations, and establishes a framework for exploring the coupling between multiple degrees of freedom in metastable laser-driven phases. These results pave the way for extending this methodology to other strongly correlated systems, enabling controlled manipulation of quantum matter beyond equilibrium.

\vspace{-0.1in}
\section{Methods}
{\bf Film preparation and characterization.} The LCMO thin films were grown on the NdGaO$_3$
substrates with the [100] or [001] crystallographic direction normal to the surface.
Using a KrF excimer laser with a wavelength of 248\,nm, LCMO films were deposited via a customized pulsed laser
deposition system at an energy density of $\sim$\,2\,J/cm$^2$ and a repetition rate of 5 Hz. The sample-to-target distance
was fixed to 5.5\,cm. During deposition, the substrate temperature and oxygen pressure were kept at 735$^\circ C$ and
40 Pa, respectively. Before cooling down to room temperature, all LCMO films were in situ annealed for 15 mins, and exhibit the ferromagnetic metallic ground state, which are labeled as the as-grown film. Then, some of LCMO films were ex situ annealed at 800 °C in flowing O$_2$ gas for 30 h, and show the antiferromagnetic insulating ground state, which are denoted as the annealed film. 
The LCMO films were structurally characterized with 2$\theta$-$\omega$ linear scans by X-ray diffraction using Cu K$\alpha$1 radiation (Panalytical X´pert) (see Supplementary Information \cite{Supplement}, Fig.~S1). The surface morphology was characterized using an atomic force microscopy (Vecco, MultiMode V). The magnetization measurements were performed using a Quantum Design magnetic property measurement system (MPMS3). The resistivity was determined by standard four-point-probe method in a temperature range of 10-300 K using a Quantum Design physical property measurement system (PPMS). The X-ray diffraction, electric resistivity, and magnetization data are in the Supplementary Information \cite{Supplement}.

{\bf Laser - transport measurements in LCMO - AFI films.} We deposited a series of electrodes on our LCMO/NGO (100/001) films with 500 or 300 $\mu$m separation distance. The voltage between the top (labeled as 1) and bottom (labeled as 2) electrodes shown in Fig. \ref{Structures}(b) is measured at the fixed current $I$ = 10 nA. The low applied current is to minimize Joule heating. The resistance is calculated using the Ohm’s law. 
The ultrafast laser spot is focused to roughly 120 $\mu$m and is guided to the center between the electrodes which can be visualized by an infrared camera. In this case, the resistance between electrodes 1 and 2 will be a combination of the un-switched and switched area since the separation distance is much bigger than the laser spot size. We therefore only focus on the percentage of changes in resistance rather than resistivity since the precise resistivity change after photoexcitation depends on how domains are reformed and connected.

{\bf Ultrafast laser setup and laser beam characterization.} We have developed an experimental setup where an ultrafast laser beam is introduced into the experimental RIXS sample chamber [see Supplementary Information \cite{Supplement}, Fig.~S12(a)]. Specifically, we use an ultrafast (1030 nm center wavelength) Yb fiber laser providing pulses width of 250 fs that is focused to $\sim$ 200 $\mu$m diameter onto the sample using a series of focusing lenses and a Keplerian-type beam expander [see Supplementary Information \cite{Supplement}, Fig.~S12(c)]. The laser beam profile is displayed in Supplementary Fig.~S12(d) \cite{Supplement}, confirming the high quality Gaussian-shape of our beam. The optics is placed out of vacuum for simplicity, and a continuous-wave (CW) alignment laser (633 nm) is integrated into the same optical path of the ultrafast laser beam by means of a beam splitter [see Supplementary Information \cite{Supplement}, Fig.~S12(c)]. In this case, the ultrafast laser spot will retrace the beam path of the CW laser and shine at roughly the same position on the sample. A camera sensitive to infrared light is used for fine tuning the location of the laser beam on the sample and align it to the x‐rays beam [see Supplementary Information \cite{Supplement}, Fig.~S12(b)]. The x-ray beam size is around 2$\times$20 $\mu$m, which is 10 times smaller compared to the laser spot. In this case, we confirm our measured area is fully irradiated by the laser and any photo-induced effect is real. The laser average power is measured by a power meter with an attached thermal power sensor. The laser power can be controlled by varying the transmission percentage of the oscillator of the laser. The maximum average power at 1 kHz is about 7000 $\mu$W [see Supplementary Information \cite{Supplement}, Fig.~S12(e)] and the beam size is $\approx$ 200 $\mu$m, which gives a maximum laser fluence of 34.9 mJ/cm$^{2}$ at an incident angle of 40 degrees [see Supplementary Information \cite{Supplement}, Fig.~S12(f)]. The fluence values quoted here include the projection factor associated with the incident angle. The fluence is varied under different laser power. We also measured the average laser power at different repetition rate from 1 Hz to 500 kHz, confirming the laser pulse energy does not vary much when changing the repetition rate.

{\bf RIXS and XAS measurements.} RIXS and XAS at the O-$K$ and Mn $L_3$ edges were performed at the 2ID-SIX beamline at NSLS-II, Brookhaven National Laboratory (USA).
The sample is mounted in a manipulator with four degrees of freedom (x, y, z, and $\theta$) [see Supplementary Information \cite{Supplement}, Fig.~S12(c)]. To overlap the ultrafast laser with the x-ray beam spot, we use the grazing-out geometry [Fig. \ref{Structures} and see Supplementary Information \cite{Supplement}, Fig.~S12(c)], in which the grazing angle $\theta$ is fixed at 140 degree to maximize the laser fluence. The scattering angle (2$\Theta$) is fixed at 150 degree to maximize the momentum transfer [see Supplementary Information \cite{Supplement}, Fig.~S12(c)]. The corresponding wavevector of the polaron is $Q$ $\approx$ [0.095, 0.095, 0.289]. For the momentum dependence measurements on polaron [see Supplementary Information \cite{Supplement}, Fig.~S4(a)], we use the grazing-in geometry with $\sigma$ polarization of the incident x-ray. Since the film is three-dimensional, we measure the polaron dispersion under the specular condition by varying $\theta$ and 2$\Theta$ correspondingly ($\theta$ = $\frac{2\Theta}{2}$) to allow the momentum transfer along the [$h$,$h$,$h$] direction (pseudo cubic notation). The XAS
spectra were collected in total fluorescence yield (TFY), using both vertical $\sigma$. The $\sigma$ and $\pi$ polarizations are defined in Fig. \ref{Structures}. All the XAS and RIXS spectra displayed in this paper are collected using the $\sigma$ polarization. The RIXS energy resolution is about 27 meV at the O-$K$ edge, as determined by elastic
measurements of a multilayer reference sample. An Aluminum filter is used to remove any infrared photons on the RIXS detector. The Fitting of RIXS spectra is detailed in the Supplementary Information \cite{Supplement}. For better visualization in Figs. \ref{RIXS with laser} and Supplementary Figs.~S2, S7, S8 \cite{Supplement}, spectra were binned by a two-point adjacent average (effective width $\approx$ 23 meV $<$ instrumental resolution $\approx$ 27 meV). All fits were cross-checked on raw and binned data and yielded consistent parameters within the experimental uncertainty (see Supplementary Information \cite{Supplement}, Fig.~S10).

{\bf First-principles calculations.} First-principles density-functional theory (DFT) calculations were performed using the Vienna $ab$ initio Simulation Package (VASP) \cite{Kresse96a} within the projector-augmented wave (PAW) framework~\cite{Blochl94}.  
The revised generalized-gradient approximation as parameterized by Perdew, Burke, and Ernzerhof (PBEsol) was employed for exchange-correlation functional~\cite{PBEsol}. 
Ten ($3s^{2}3p^{6}4s^{2}$), eleven ($5s^{2}5p^{6}5d^{1}6s^{2}$), thirteen ($3p^{6}3d^{5}4s^{2}$), and six ($2s^{2}2p^{4}$) valence electrons were considered in the PAW pseudopotentials of Ca , La, Mn, and O, respectively.
Strong correlation effects for Mn 3$d$ electrons were treated at the mean-field level using the rotationally invariant DFT+$U$ method introduced by Liechtenstein et al.~\cite{Liechtenstein}. 
We set $U$(Mn-3$d$) = 4\,eV, $J$(Mn-3$d$) = 1\,eV, $U$(La-4$f$) = 6\,eV, and $J$(La-4$f$) = 0\,eV following Refs.~\cite{Hong2020}. 
We find that this set of values appropriately describes the lattice parameters, magnetic structure, and vibrational properties of La$_{2/3}$Ca$_{1/3}$MnO$_3$. 
La/Ca concentration ratio of 0.67/0.33 was simulated using the virtual-crystal approximation~\cite{Bellaiche2000} which has been reported to produce reliable results for similar systems~\cite{Trappen2018}. 
The total energy convergence and residual force convergence criteria were set to $10^{-7}$\,eV and $10^{-3}$\,eV/\AA, respectively. The cutoff energies for the plane-wave basis set was set at 600\,eV. We used a Monkhorst-pack $k$-mesh~\cite{MP1976} of size 12$\times$12$\times$9 to sample the reciprocal space. Electronic structure was analyzed using the {\sc Pyprocar} package~\cite{herath2020}.

\section{Acknowledgements}  
We thank Dr.~John Freeland, Prof.~Andrew Millis, Dr. Mark Dean, and Prof.~Jingdi Zhang for useful discussions. Work at Brookhaven National Laboratory was supported by the DOE Office of Science under Contract No.~DE-SC0012704. This work was supported by the Laboratory Directed Research and Development project of  Brookhaven National Laboratory No.~21-037. UK was supported by the US Department of Energy, Office of Basic Energy. This work was supported by the U.S.~Department of Energy (DOE) Office of Science, Early Career Research Program. This research used Beamline 2-ID of the National Synchrotron Light Source II, a U.S.~Department of Energy (DOE) Office of Science User Facility operated for the DOE Office of Science by Brookhaven National Laboratory under Contract No.~DE-SC0012704. The authors at the University of Science and Technology of China acknowledge the financial support from Quantum Science and Technology Major Project, with Grant No. 2024ZD0301300. The authors also acknowledge the Anhui Provincial Natural Science Foundation with Grant No. 2308085MA15. SS was supported by the U.S.~Department of Energy, Office of Science, Office of Fusion Energy Sciences, Quantum Information Science program under Award Number DE-SC-0020340. SS also acknowledges support from the University Research Awards at the University of Rochester. O.S.B. and A.M. acknowledge financial support from the Croatian Science Foundation, Grants No. HRZZ-IP-2022-10-3382 and HRZZ-IP-2024-05-2406, respectively.

\section{\bf Competing Interests}
The authors declare no competing interests.

\section{\bf Data availability}
\vspace{-0.2cm}
Relevant data are available upon reasonable request from the corresponding authors.


%

\end{document}